\documentclass[printer]{aa}
\usepackage{epsfig}
\usepackage{graphicx}

\begin{document}
\title{Stellar density distribution in the NIR on the Galactic plane
at longitudes 15-27 deg.\\
Clues for the Galactic bar ?}
\author{S. Picaud\inst{1}, A. Cabrera-Lavers\inst{2} \and F. Garz\'on\inst{2,3}}
\offprints{S. Picaud}
\mail{picaud@obs-besancon.fr}
\institute{CNRS UMR6091, Observatoire de Besan{\c c}on, BP 1615,
F-25010 Besan{\c c}on Cedex, France
	\and
Instituto de Astrof\'{\i}sica de Canarias, 38200 La Laguna, Tenerife, Spain
	\and
Departamento de Astrof\'{\i}sica, Universidad de La Laguna, Tenerife, Spain}
\authorrunning{S. Picaud et al.}
\titlerunning{Stellar density at longitudes 15-27 deg.}

\date{Received 6 March 2003 / Accepted 28 May 2003}

\abstract{Garz\'on et al. (1997), L\'opez-Corredoira et al. (1999) and
Hammersley et al. (2000)
have identified in TMGS and DENIS data a large excess of stars
at l=27$^{\circ}$ and b=0$^\circ$ which might correspond to an in-plane
bar. We compared near infrared CAIN star counts and simulations from the
Besan\c{c}on Model of
Galaxy on 15 fields between 15$^\circ$ and 45$^\circ$ in longitude and
-2$^\circ$ and 2$^\circ$ in latitude. Comparisons confirm the existence of an
overdensity at longitudes lower than 27$^\circ$ which is inhomogeneous and
decreases very steeply off the Galactic plane.
The observed excess in the star distribution over the predicted density
is even higher than 100\%. Its distance from the sun is estimated to
be lower than 6 kpc.
If this overdensity corresponds to the stellar population of
the bar, we estimate
its half-length to 3.9$\pm$0.4 kpc and its angle from the Sun-center
direction to 45$\pm$9 degrees.
}
\maketitle

\section{Introduction}

Star counts have been used for years to examine at various levels of
detail the stellar contents in the Galaxy (see Paul 1993), whose structural
parameters of the various morphological components are still far from being
completely known. In the last two decades there has been a combined effort in
this direction with the use of detailed models of stellar galactic distribution
(Bahcall \& Soneira 1980; Robin \& Cr\'ez\'e 1986; Wainscoat et al. 1992;
L\'opez--Corredoira et al. 2002)
along with large area, high sensitivity and multicolour star count
surveys. The near infrared (NIR) members of these  surveys (Eaton et
al. 1984, Garz\'on et al.
1993, Hammersley et al. 1994, L\'opez--Corredoira et al. 2000, Epchtein 1997,
Skrutskie et al. 1997) are notably useful for the analysis of the
galactic structure because of less interstellar extinction compared to the
optical bands, in particular in the hidden on-plane areas of the inner Galaxy.
It is in these zones that the morphological structures are less well studied.

The effect of the high extinction towards the inner Galaxy along the Galactic
plane, added to the position of the Sun very near to the plane itself
(Hammersley et al. 1995), makes it particularly difficult to detect large
scale stellar structures in the central part of the Galaxy.
There is an increasing
consensus that the Galaxy is of barred type (see Garz\'on 1999 for a review),
but there is still a considerable controversy about the morphology and stellar
content of the bar.

Hammersley et al. (2000) has recently derived
arguments for the Galactic bar to be of half-length of
roughly 4 kpc and with a position angle with the sun-center direction
of around $43\deg$, by observing from
new NIR star count data an excess of stars at longitudes on the Galactic plane
at l$\leq$27$^\circ$ and assuming them to belong to the bar population.
To continue this research work (see also L\'opez--Corredoira
et al. 2001) in the field of the structure of
the inner Galaxy we have made use in this paper of the large NIR star count
database obtained with the CAIN camera at the Teide Observatory (Hammersley et
al. 2000) together with the Besan\c{c}on model of the Galaxy (Robin
\& Cr\'ez\'e 1986; Bienaym\'e et al. 1987; Robin et al. 2003)
to examine the
excesses in both the magnitude and the colour histograms of the data with
respect to the model predictions.
This study permits us on the
one hand to confirm with an independent method the existence of this extra
density,
and allows us on the other hand to derive considerations about its structure.
Indeed,  since the Besan\c{c}on model does not contain
any galactic component other than the thin disc (and possibly the triaxial
bulge at the innermost field) in the regions of
interest, as will be described in the next sections, any excess observed in
the
stellar density compared to the model prediction may then be used to analyze
the morphological extent of such an extra population.

This paper is organized as follows. We briefly describe the star count
data and the Besan\c{c}on model used in this work. We then discuss the
method of deriving the extinction distribution along the line of sight from the
data alone, which is then used in the model for the predictions.Those
predictions are then compared with the data in two different planes:
magnitude and colour counts, which both serve for the analysis.
We provide
conclusions describing the model itself and the derived geometry for the
component responsible for the excesses.

\section{The Besan\c{c}on Model of the Galaxy}

The approach of the Besan\c{c}on model is slightly different from the
ones used previously to study the bar. Here,we give
a brief description of the model, dealing particularly with the thin disc
population model. More details can be found in Robin \& Cr\'ez\'e (1986),
Bienaym\'e et al. (1987) and Robin et al. (2003)\footnote{A version of the
model is available on  the web at \\
http://www.obs\-besancon.fr/fesg/modele\_ang.html}.

\subsection{Global description}

The Besan\c{c}on Model of Stellar Population Synthesis,
developed in the visible, near and mid infrared, aims at giving a
global 3-dimensional description of the Milky Way, including stellar
populations such as the thin disc,
outer bulge, thick disc and spheroid, as well as dark halo
and interstellar matter. 

Its approach is semi-empirical: both theoretical
schemes (stellar evolution, galactic evolution, galactic dynamics) and
empirical laws found in the literature are used.
Boltzmann and Poisson equations
are used to make the model dynamically self-consistent in the direction
perpendicular to the plane. Artificial catalogues of stars are produced,
giving both the fundamental characteristics of stars
(age, distance, absolute magnitudes,
kinematics) and observable ones (apparent magnitudes, colours, proper motion,
radial velocities) directly comparable with observations. Absorption,
photometric errors and Poisson noise are also added to make simulations 
as close as possible to observations.

The Besan\c{c}on model has been used to constrain the density law or the
luminosity function of the stellar populations: thin disc (Ruphy et al. 1996;
Haywood et al. 1997), thick disc (Robin et al. 1996; Reyl\'e \& Robin 2001),
spheroid (Robin et al. 2000) and bulge (Picaud \& Robin, in preparation).

\subsection{Thin disc}

In the studied region, thick disc and spheroid populations give a
negligible contribution, and those from the bulge likely only contaminate
star counts at
the innermost field (l=15$^\circ$). Thus, the only dominant stellar population
taken into account by the model is the thin disc. Its density is modeled as
the Einasto law (1979): the thin disc is divided in 7 age components and
each component
distribution is described by an axisymmetric ellipsoid. The first
component (0-0.15 Gyr) is called the
\emph{young disc} and the 6 other ones form
the \emph{old disc}, which only is presented here because of the very small
number of stars belonging to the young one. The density law of each
ellipsoid of the old disc is the subtraction of two modified exponentials,
the second one representing the central hole:

$$\rho = \rho_0 \times [\exp (-\sqrt{0.25+ ( \frac{a}{R_d} )^2} )
-\exp (-\sqrt{0.25+ ( \frac{a}{R_h} ) ^2} ) ]$$
\begin{center}
with $a^2=R^2+\left(\frac{Z}{\epsilon}\right)^2$, where:
\end{center}

\begin{itemize}

\item $R$ and $Z$ are the cylindric coordinates

\item $\epsilon$ is the axis ratio of the ellipsoid which increases with the
age of the component. The 6 axis ratios of the old thin disc are presented in
Table \ref{axisratios}.

\item	$R_d$ is the global scale length of the disc and 
$R_h$ is the global scale length of the hole.
In the present version of the Besan\c{c}on model, we use $R_d$=2.37 kpc
and $R_h$=1.31 kpc,
deduced from model fitting toward the inner Galaxy (Picaud \& Robin, in
preparation).

\item The normalization $\rho_0$ was deduced from the local luminosity
function (Jaheiss et al, private communication).
\end{itemize}

 \begin{table}
{\centering
\begin{tabular}{lc}
\hline
\hline
\textbf{age (Gyr)}& $\epsilon$ \\
\hline
0.15-1 	& 0.0268\\
1-2	& 0.0375\\
2-3	& 0.0551\\
3-5	& 0.0696\\
5-7	& 0.0785\\
7-10	& 0.0791\\
\hline
\label{axisratios}
\end{tabular}
\caption{Axis ratios of the 6 age components of the old thin disc.}\par}
\end{table}

The distributions in $M_v$, $\log{T_{eff}}$, $\log{g}$ are deduced
from an evolution model described in Haywood
et al. (1997). Absolute magnitudes in K are computed from the corresponding V
absolute magnitudes using the semi-empirical model of atmospheres from Lejeune
et al. (1997, 1998). Disc density parameters and further changes in the
luminosity function are explained in Robin et al. (2003).

\subsection{Extinction}

The Besan\c{c}on model includes its own extinction model which follows the
young thin disc distribution, with a scale length of 4 kpc, an equivalent scale
height of 140 pc, and a normalization which can be modulated. However,
this model can be replaced by another distribution as needed.

For this study, we have preferred to model the extinction without using the
Besan\c{c}on simulations to avoid a possible bias in the comparisons of the
model with the data. The determination of the extinction distribution
in each field will be described in section \ref{extinction}.

\section{Observations}

Since 1999 the IAC group has been building a Galactic survey by using CAIN, the
NIR camera on the 1.5-m Telescopio Carlos S\'anchez (TCS) (Observatorio
del Teide, Tenerife). Observations of a series of 20 x 12 arcmin$^{2}$
(i.e., a total area of approximately 0.07 deg$^2$) have been made along the
Galactic plane between $l=0^{\circ}$ and $l=220^{\circ}$,
together with some off-plane
series (mainly in b=$\pm2^\circ,\pm5^\circ$ and $\pm10^\circ$). The survey uses
the J (1.25$\mu$m), H (1.65$\mu$m), and K$_{short}$ (2.17$\mu$m) standard
bands,
with limiting magnitudes of 17, 16.5 and 15.2 respectively. This provides
almost one magnitude deeper coverage
than 2MASS or DENIS surveys in the K-band (Struskie
et al. 1997; Epchtein et al. 1997), which is very useful for the study of the
inner
galactic regions such as those considered in this paper. For the fields we used
here, the seeing was typically of 1 arcsec, and data were obtained only in
photometric conditions. The main characteristics of the 15 chosen fields and
photometric bands used are summarized in Table \ref{tabcamps}.

\begin{table*}
{\centering
\begin{tabular}{lcccc}
\hline
\hline
\textbf{l (deg)}&
\textbf{b (deg)}&
\textbf{area (x 10$^{-2}$ deg\( ^{2}) \)}&
\multicolumn{2}{c}{\textbf{completeness limits (band used)}}\\
\hline
15 & 0 & 6.95 & 16.8 ($J$) & 14.8 ($K_s$)\\
20 & 0 & 6.97& 16.8 ($J$) & 14.8 ($K_s$)\\
21 & 0 & 6.99& 16.8 ($J$) & 14.8 ($K_s$)\\
26 & 0 & 12& 16.8 ($J$) & 14.6 ($K_s$)\\
26 & 2& 6.95& 16.4 ($J$) & 14.8 ($K_s$)\\
27 & 0& 6.97& 16.4 ($J$) & 14.4 ($K_s$)\\
27 & 0.5 & 6.97& 16.4 ($J$) & 14.8 ($K_s$)\\
27 & -0.5& 6.96& 16.4 ($J$) & 14.8 ($K_s$)\\
27& 1& 6.91& 17 ($J$) & 15 ($K_s$)\\
28& 0 & 6.93& 17 ($J$) & 15 ($K_s$)\\
32& 0 & 6.92& 17 ($J$) & 15 ($K_s$)\\
33& -2& 7.03& 16.4 ($J$) & 14.8 ($K_s$)\\
37& 2& 7.03& 16.8 ($J$) & 15.4 ($H$)\\
40& 0 & 7.05& 16.8 ($J$) & 14.8 ($K_s$)\\
45& 0 & 6.99& 16.6 ($J$) & 14.6 ($K_s$)\\
\hline
\label{tabcamps}
\end{tabular}
\caption{CAIN survey selected regions used in this paper.}\par}
\end{table*}

\section{Extinction from K-giants}\label{extinction}

Extinction changes very much from one field to another in and close to
the Galactic plane, and a good determination of its distribution is needed
to compare data and simulations in this region. This is why a global extinction
model is not sufficient and the distribution of extinction along the line of
sight must be determined field by field.

\subsection{Method of extraction of the extinction distribution}

L\'opez-Corredoira et al. (2002) (hereafter L02) developed a method to
obtain the star density and interstellar extinction along a line of sight by
extracting a well-known population (spectral type K2III) from the infrared
colour-magnitude diagrams (hereafter CMD). The method is extensively
described in L02, so we give only a brief summary here.

($J-K,m_K$) CMDs are built for each field. In the diagrams, stars of the same
spectral type (which would mean that they have about the same absolute
magnitude but are located at different distances from the Sun)
will be placed at different locations in the CMD. The effect of distance alone
shifts the stars vertically, while extinction by itself shifts the stars both
horizontally and vertically. Combinations of these effects, plus some
intrinsic dispersion in absolute magnitude and/or spectral type, cause
the red clump giant stars,
which constitute the majority of the disc giants (Cohen et al. 2000; Hammersley
et al. 2000), to form a diagonal broad branch running from top left to bottom
right on the CMDs.

In order to isolate the red clump sources, count histograms have been made by
making horizontal cuts (i.e. running colour) through the CMDs at fixed $m_{K}$.
A Gaussian function was then fitted to the histogram to determine the position
of the peak in each cut.

\begin{figure}
\resizebox{\hsize}{!}{\epsfig{file=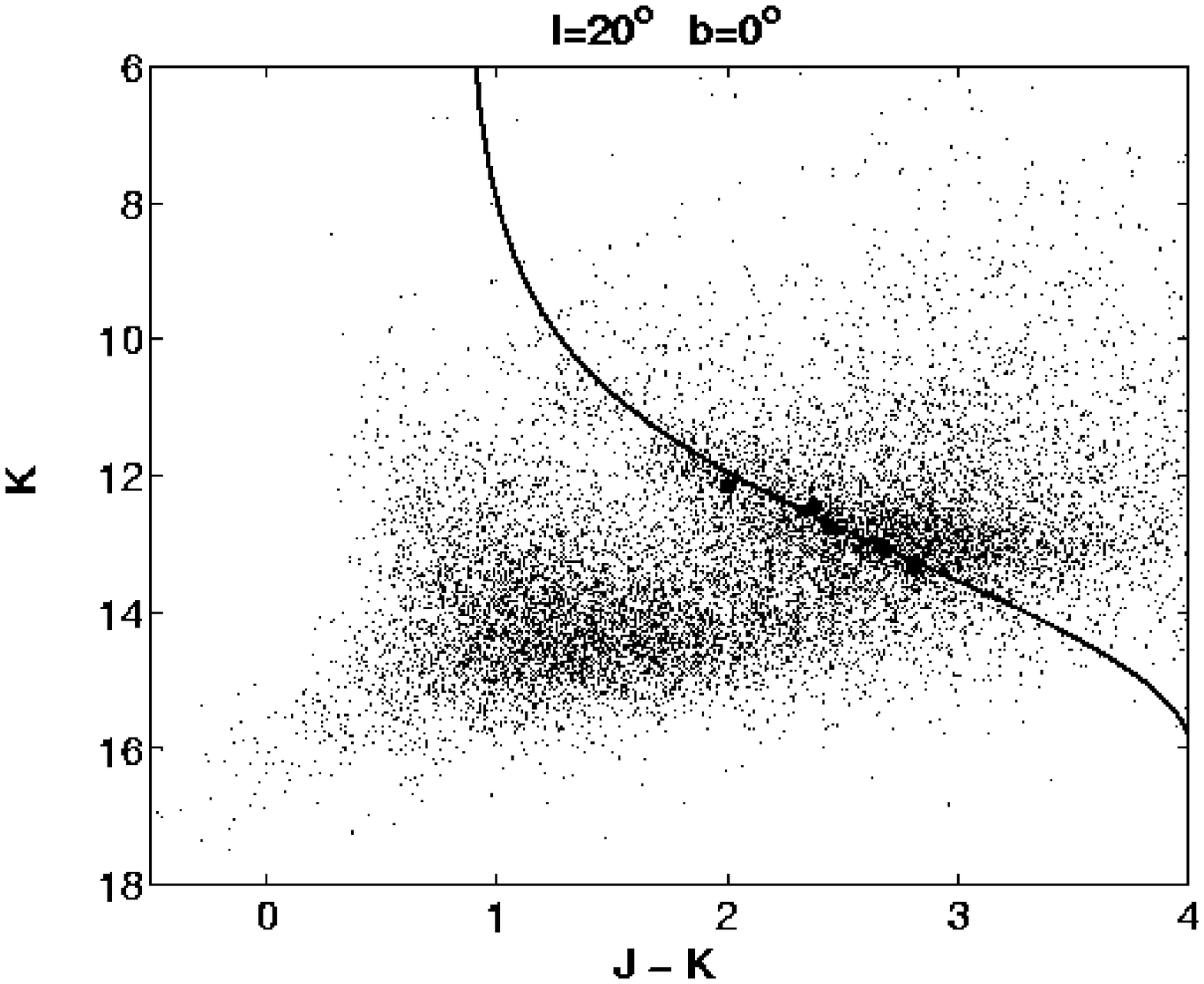}
\epsfig{file=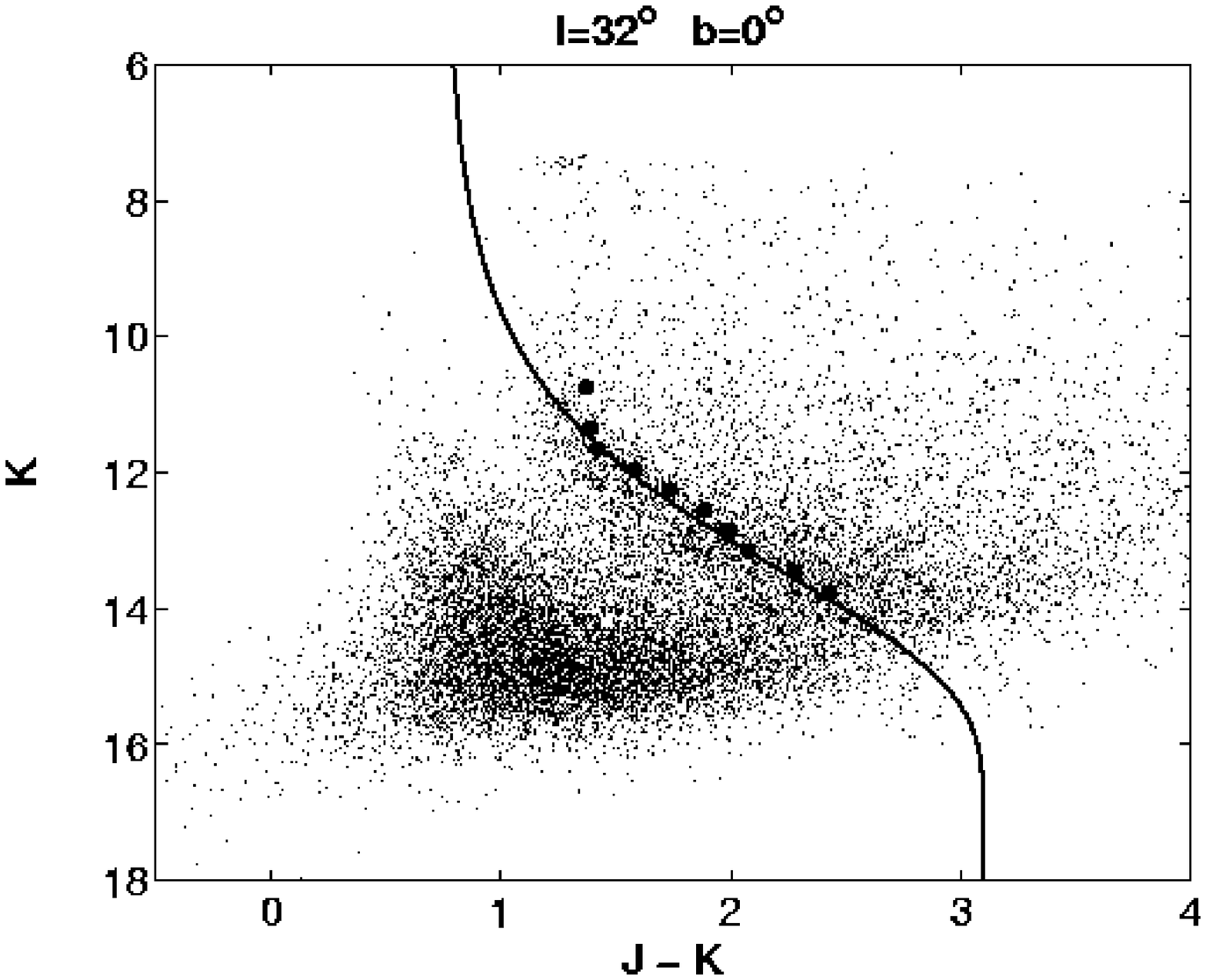}}
\caption{CMDs for two of the fields studied here. Maxima obtained by a
Gaussian fitting to star distribution in the red clump strip are
over-plotted. The solid line shows the fitted trace assigned to the red
clump population affected by varying extinction along the line of sight.}
\label{Fig:CM1}
\end{figure}

The extinction A$_{K}(m_{K})$, to a distance
$r$, can be determined by tracing how the (J-K) of the peak of the red clump
counts changes with $m_{K}$. The extinction is calculated for any given $m_{K}$
using the measured $(J-K)$ of the peak, the intrinsic mean colour excess
definition, and the interstellar extinction values for
$\frac{A_{J}}{A_{V}}$ and $\frac{A_{K}}{A_{V}}$ given by
Mathis (1990).

\begin{equation}
A_{K}=\frac{(J-K)-(J-K)_{0}}{1.61}
\label{ak}
\end{equation}

Finally, the extinction law ($A_{K}$ vs. $r$) can be built for each field by
using a suitable transformation for the distance along the line of sight:

\begin{equation}
r=10^{\frac{m_{K} - M_K + 5 - A_{K}(r)}{5}}
\label{r}
\end{equation}

Uncertainties in the method, such the metallicity effects for the clump
giants, the effect of the dwarf contamination in the K-giant strip, or
differences between the real red clump distribution and the Gaussian function
are fully described in L02.

\subsection{Absolute magnitude calibration}
\label{BM}

\begin{figure}
\resizebox{\hsize}{!}{\epsfig{file=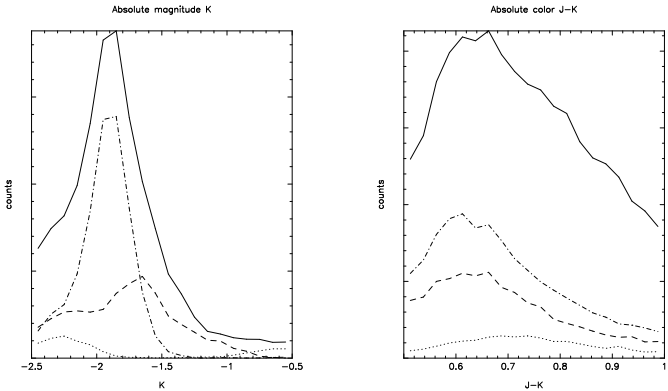,angle=270}}
\caption{Distribution of K absolute magnitudes and $(J-K)_0$ colours according
to the Besan\c{c}on model. The K0 counts are represented with 
dotted lines, the K1 with dotted-dashed lines and the K2 with dashed lines,
while the continuous line represents the sum of the three counts.
The maxima of the K1 plots correspond to the red
clump, with an absolute magnitude of -1.85 and an intrinsic colour of 0.64.}
\label{Fig:BMod}
\end{figure}

The method described above can provide spatial information from the CMDs, and
the only assumption being made is that the absolute magnitudes of all the
sources being extracted is more or less fixed. In order to deduce
the distribution of extinction along the line of sight, absolute magnitudes
and colors of stars must be assigned.
In L02, the peaks were identified as pertaining to the K2III
stars since they are by far the most prominent population, according to
the updated ''SKY'' model (Wainscoat et al. 1992; M. Cohen, private
communication)(see Fig.2 in L02 for details).
The K2III population mean absolute magnitude and the intrinsic color were
assumed to be $M_{K}=-1.65$ and $(J-K)_{0}$=0.75, with a gaussian dispersion
of 0.3 mag in absolute magnitude and 0.2 in colour.
However, in the present study, the absolute magnitude and intrinsic color
values must be the ones used in the Besan\c{c}on model to get a consistent
analysis of the data.
The Besan\c{c}on model predicts a very prominent population of K1 giants with
an absolute magnitude of $M_K$=-1.85 and an intrinsic colour $(J-K)_0$=0.64
(Fig.\ref{Fig:BMod}). These values are slightly different from those considered
initially in L02. The determination of spectral types in the
model is taken from De Jager \& Nieuwenhuijzen (1987).

\begin{figure}
\resizebox{\hsize}{!}{\epsfig{file=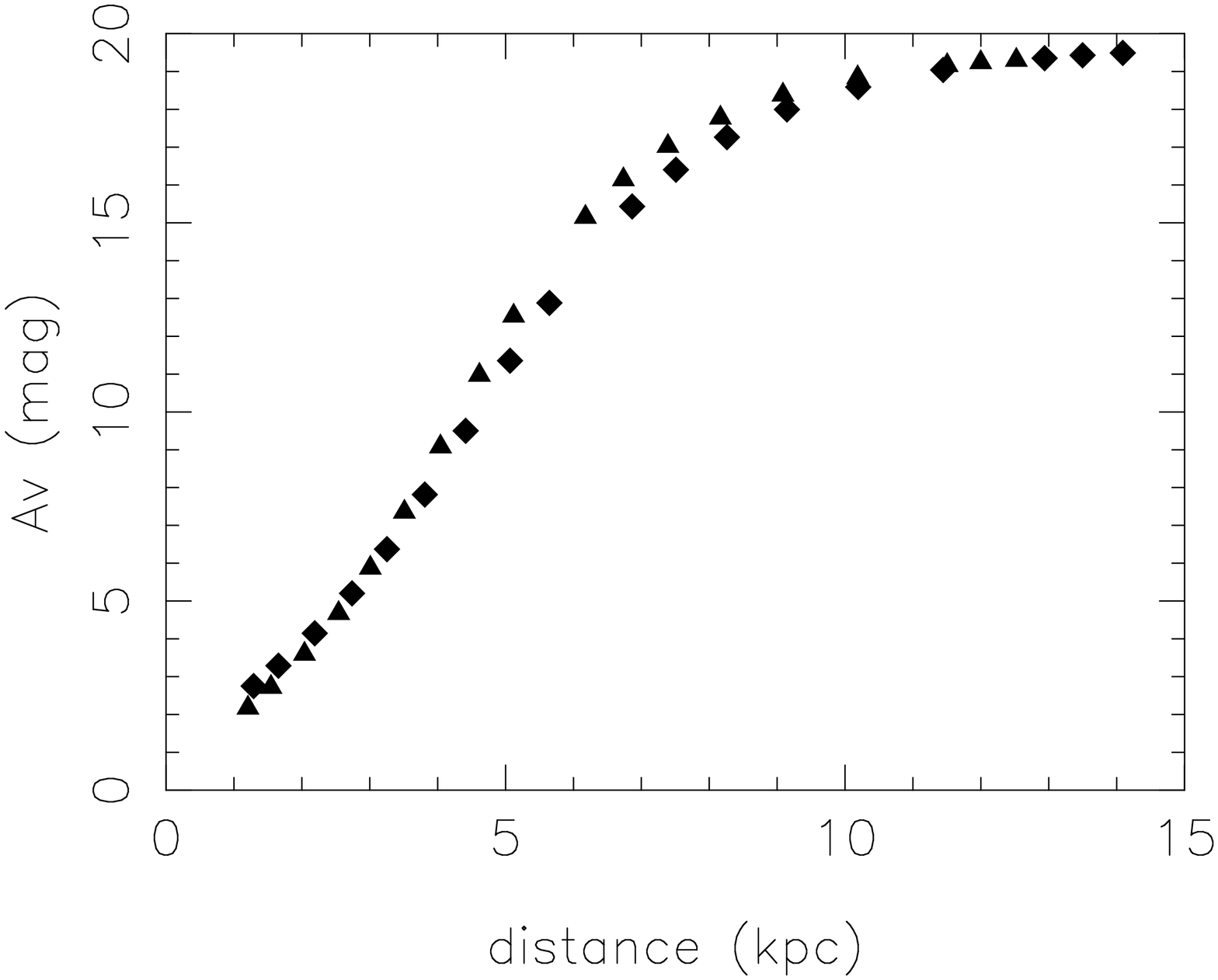,height=6.7cm,angle=-90}
\epsfig{file=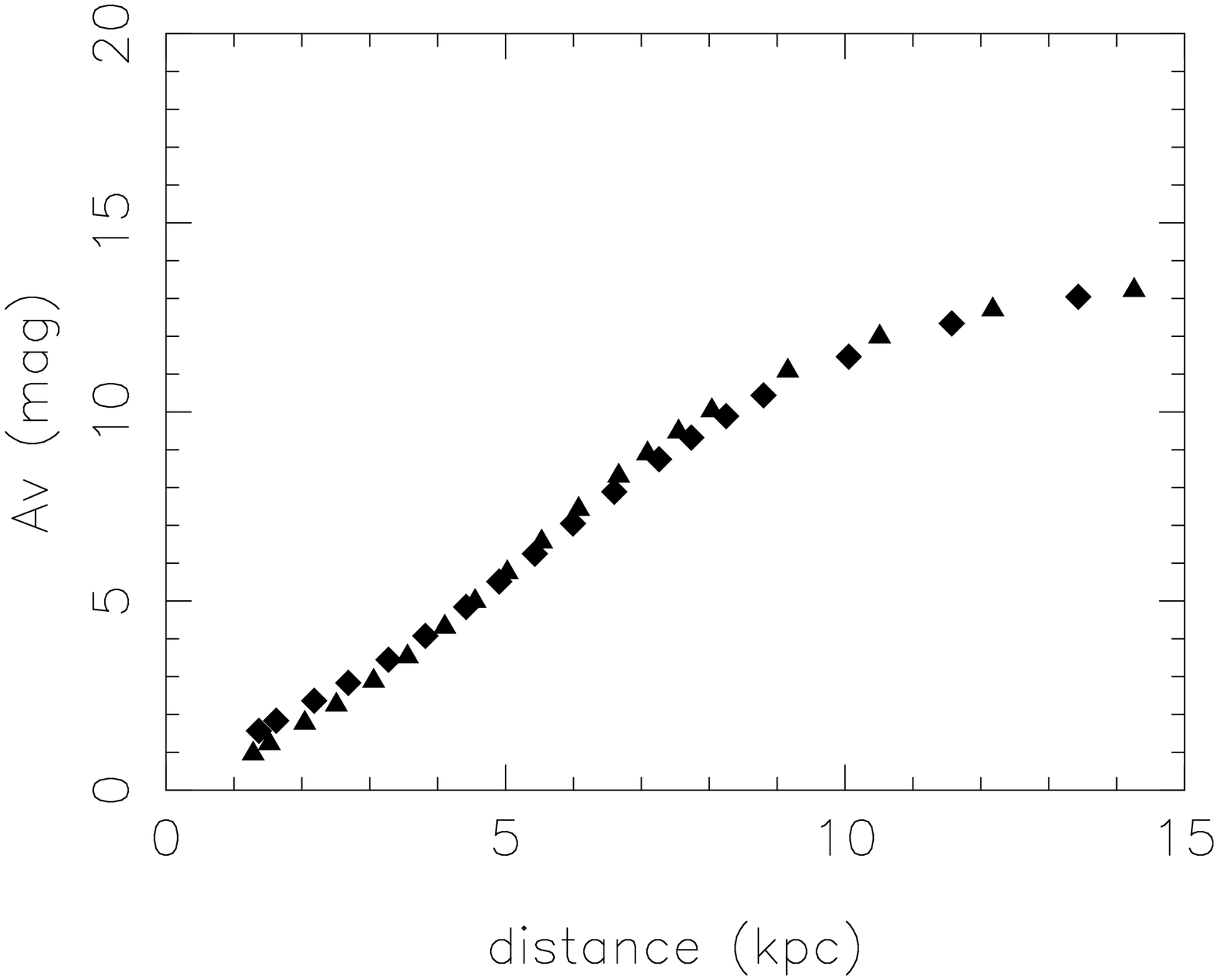,height=6.7cm,angle=-90}}

\caption{Extinction along the line of sight ($A_V$ vs $r$)
obtained for the fields l=20$^\circ$, b=0$^\circ$(left) and
l=32$^\circ$,b=0$^\circ$(right). The extinctions obtained
assuming $M_K$=-1.65 and $(J-K)_0$=0.75 (L02) are shown with triangles while
values with $M_K$=-1.85 and $(J-K)_0$=0.64 (Besan\c{c}on model) are shown with
diamonds. Differences are negligible,
and they are only noticeable for off-plane regions where the extinction is
lower, hence increasing the relative difference.}

\label{extlaw}
\end{figure}

Differences in using either the values used by L02 or those of Besan\c{c}on
are quantified and shown in Fig. \ref{extlaw}.
They cannot be a critical issue in the
method, since the extraction of the peaks for the star distribution in the
K-giant strip is independent of the subsequent analysis of the spectral type
corresponding to those peaks. The change in the absolute magnitude and
intrinsic colour of 
the population affects only the values extracted for A$_K$ and $r$.
The differences are less than 7\% in extinction and less than 9\% in
distance (see Fig.\ref{extlaw}). Those errors are also of the same order of
magnitude, or even less, than the
intrinsic uncertainties of the real values for the absolute magnitude and
intrinsic colour of the red clump population.

\subsection{H band.}

The analysis is made by using ($J-K,m_K$) CMDs for all the fields,
except one (l=$37^\circ$,b=$-2^\circ$) where the K$_{s}$
band was not available. A ($J-H,m_H$) CMD was used instead. The main
difference between using the H
band or the K$_{s}$ band is related to the absolute magnitude and the
intrinsic colour of the red clump population. Those values are not as well
determined for the H filter as for the K$_{s}$, and they present a bigger
dispersion than in K$_s$ (Wainscoat et al. 1992). As explained in \S\ref{BM},
we use the values defined in the Besan\c{c}on model for the red clump
population in this band ($M_H$=-1.65, $(J-H)_0$=0.5). The procedure of
deriving the
extinction is exactly the same as described before, but replacing $M_K$ and
$(J-K)$ by $M_H$ and $(J-H)$. We are analyzing an off-plane field,
and the extinction is low enough to produce significant differences in the
extinction distribution obtained by using either the H band or the K$_{s}$
band.

\section{Data vs. model predictions}

This study consists of comparisons of simulations from the
Besan\c{c}on model
with CAIN data at longitudes 15$^\circ$-30$^\circ$ to detect and study
the overdensity observed in previous studies (Hammersley et al. 2000,
L\'opez--Corredoira et al. 1999). In this region, model
counts are  dominated by the thin disc population. This is why first
comparisons between simulations and data were needed in disc fields in and
out of the plane to verify that the Besan\c{c}on model reproduces well the
thin disc counts. Once this verification has been made, counts are compared
on the one hand in the plane varying the longitude and on the other hand 
at different latitudes to extract information about both the horizontal
extent and the vertical thickness of the detected overdensity.

Model simulations have been done assuming photometric errors close
to those of the data and an extinction along the line of sight determined
by the method described in section \ref{extinction}.

Comparisons are made
in K$_s$ magnitude and J-K$_s$ colour for all fields except for the field
at l=37$^\circ$ b=+2$^\circ$ where the K$_s$ band was not available and was
replaced by the H band, as explained before.

The values of limits given in
Table \ref{tabcamps} show that in fields with high extinction,
data are not complete in the J band for the faintest and most reddened stars.
Star counts have then been produced until
K$_s$=13.5 (H=14 at l=37$^\circ$) and
J-K$_s$=3 to ensure completeness.
These new thresholds do not guarantee perfect completeness for all fields,
but this study needed counts deep enough to observe the overdensity.

\begin{figure}
\resizebox{\hsize}{!}{\epsfig{file=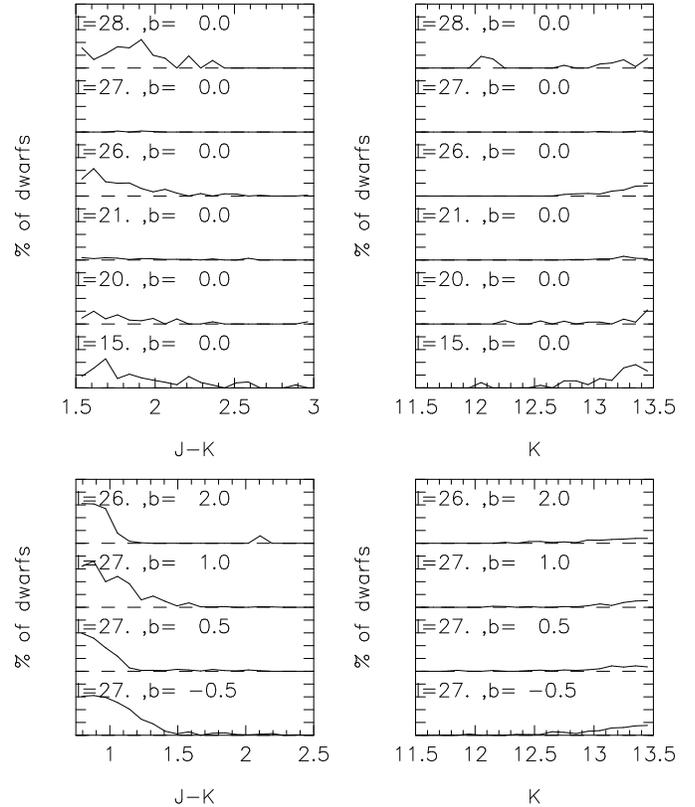}}
\caption{Percentage of dwarfs in Besan\c{c}on simulations by bins of
J-K$_s$ (left) and K$_s$ (right), for in-plane fields (up), and
off-plane ones (bottom). A step between two zero lines
(dashed lines) correspond to 50\% of dwarfs.}
\label{contamination}
\end{figure}

In order to reduce the contamination of foreground dwarfs in star
counts, only stars with J-K$_s\geq$1.5 for in-plane fields and J-K$_s\geq$0.75
for off-plane ones\footnote{The reddening being low, comparisons have
to be done at lower J-K$_s$. This implies a substantial contamination of dwarfs
for off-plane fields.} were selected. But the estimation (given by
Fig. \ref{contamination}), using the
Besan\c{c}on simulations, of the dwarf contamination 
shows that there is still up to 20\% of dwarfs
at faint K$_s$ and at low J-K$_s$ for in-plane fields, and up to
40\% at low J-K$_s$ for off-plane ones. This contamination implies
uncertainties in this work, especially for a quantitative study.

\subsection{Disc fields}

\begin{figure*}
\centering
\epsfig{file=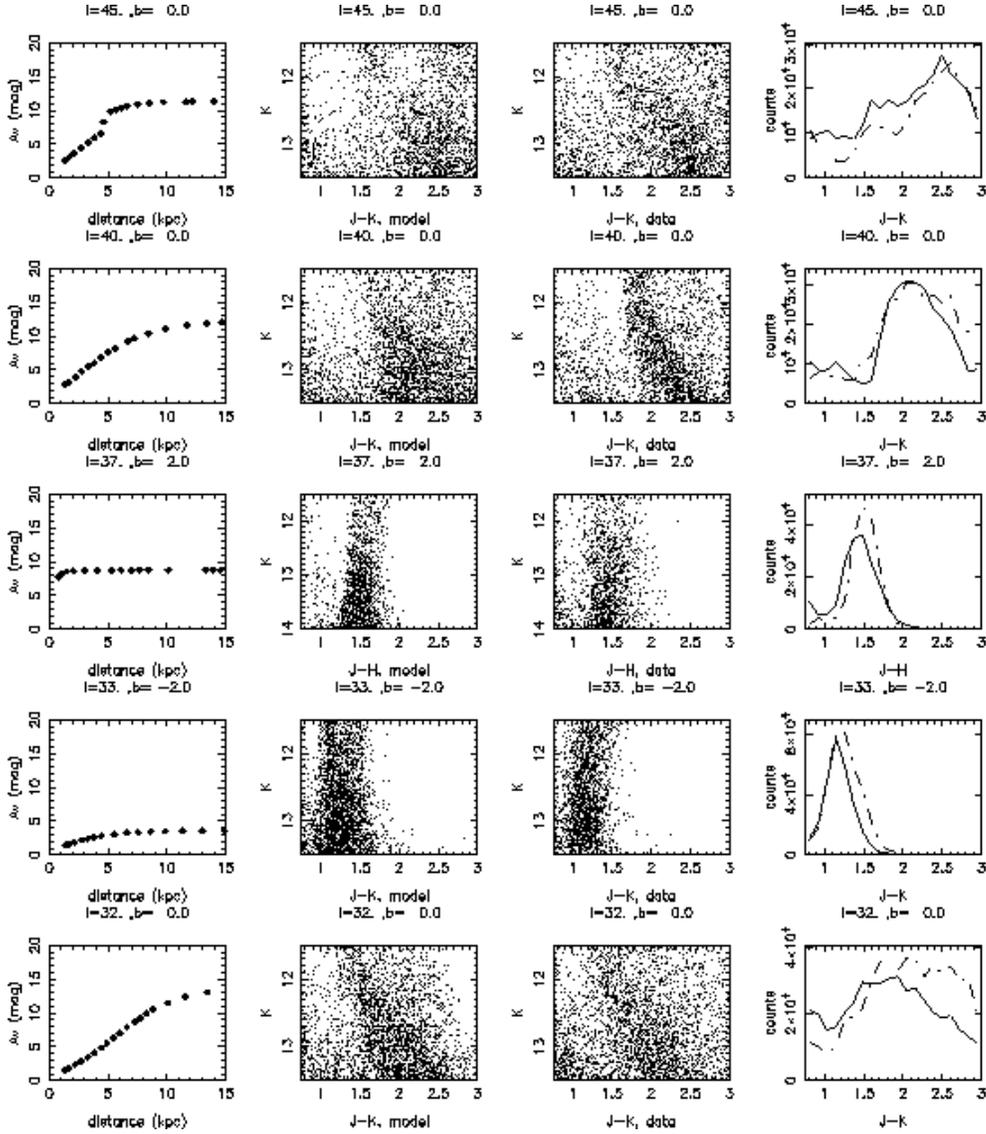,width=13cm}
\caption{Extinction distributions along the line of sight, colour magnitude
diagrams and colour histograms of intermediate disc
fields. The CMD on the left and the dashed line at the histogram panels
correspond to the Besan\c{c}on simulations, while the CMD on the right and the
solid line in the histograms display the CAIN data. Star counts of the
colour histograms (in number of stars per magnitude bin and per square
degree) are taken for 11.5 $<$ K$_s$ $<$ 13.5}.
\label{diag-disque}
\end{figure*}

Model counts were compared with data on 3 fields on the galactic plane:
b=0$^\circ$ l=45$^\circ$,
40$^\circ$ and 32$^\circ$, and 2 off plane fields: l=37$^\circ$
 b=+2$^\circ$ and l=33$^\circ$ b=-2$^\circ$. 

Extinction distributions, model and data colour-magnitude diagrams and count
histograms are presented in Figure \ref{diag-disque}.

Model and data CMDs and histograms
are rather close to each other, but in some fields there is a slight excess of
reddest stars. The model of disc stellar evolution (Haywood et al.
1997) might overestimate the duration of late type stars and perhaps the oldest
stars of these types have already become white dwarfs or planetary nebul\ae.
A correction has been made by rejecting giants with a spectral type later
than K5 and age greater than 7 Gyr but does not eliminate all the excess.
The problem of late type stars may be enhanced in off-plane fields:
in these fields, stars are less reddened and most of them
are below the cuts
in K$_s$=13.5 and J-K$_s$=3. Moreover, because of the very small reddening,
histograms of these fields may be more contaminated by foreground dwarfs even
with the blue cut in J-K$_s$.

In any case, this excess of simulated stars is small enough to consider
that the Besan\c{c}on model reproduces rather well the thin disc and to
continue the study on fields at l$<$30$^\circ$.

\subsection{Along the Galactic plane}

\begin{figure*}
\centering
\epsfig{file=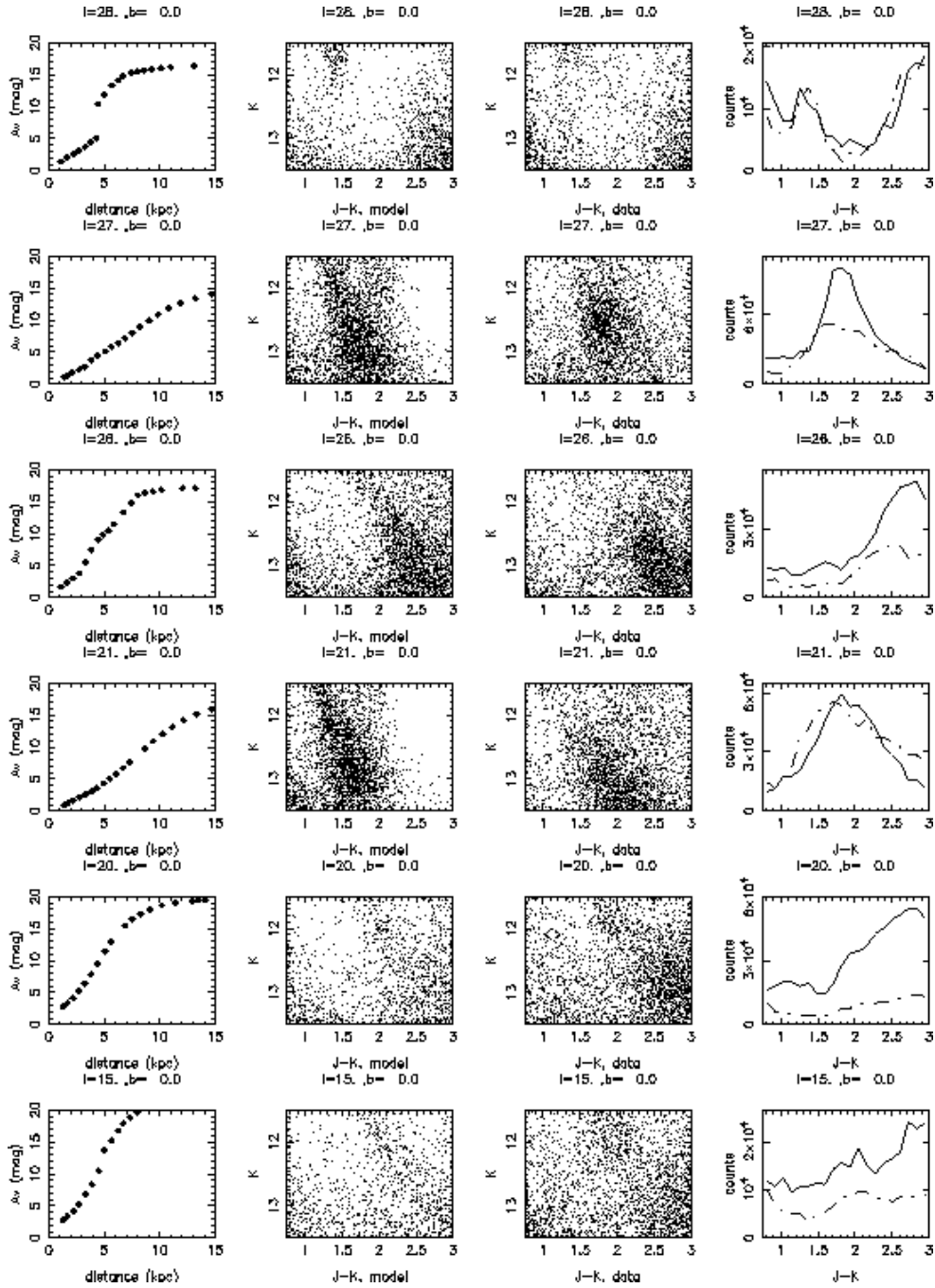,width=13cm}
\caption{Colour magnitude diagrams and colour histograms of fields at
l$\leq$28$^\circ$ and b=0$^\circ$. See caption of Figure \ref{diag-disque} for
an explanation about the content of the different panels.}
\label{diag-ldp}
\end{figure*}

In addition to the 3 on-plane disc fields (l=32$^\circ$, l=40$^\circ$,
l=45$^\circ$), we compared model and data on 6 fields at l$\leq$28$^\circ$
and b=0$^\circ$ : l=28$^\circ$, l=27$^\circ$, l=26$^\circ$, l=21$^\circ$,
l=20$^\circ$, l=15$^\circ$. Extinction distributions, colour-magnitude
diagrams and histograms are presented in Figure \ref{diag-ldp}.

\begin{figure*}
\centering
\epsfig{file=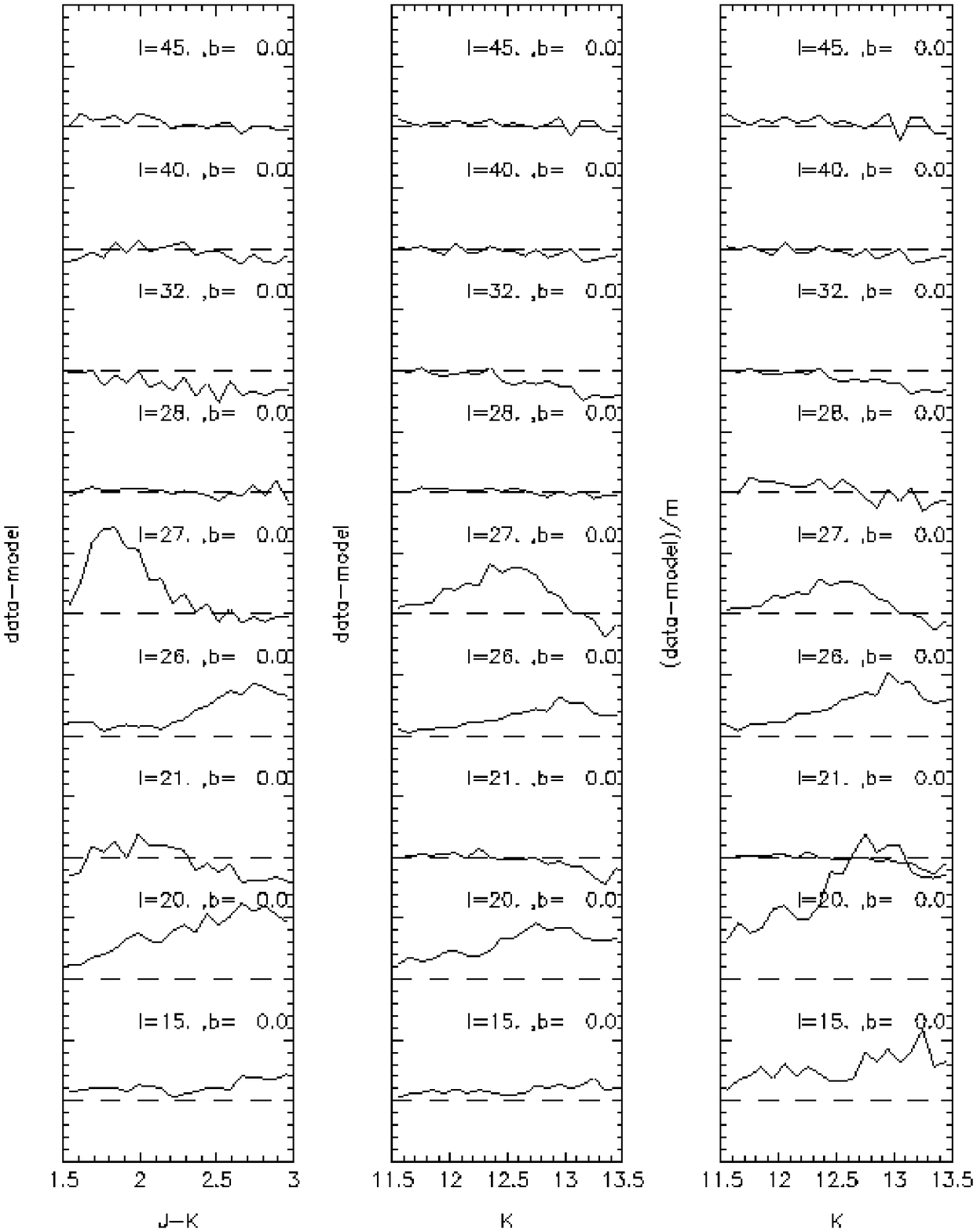,width=13cm}
\caption{Difference distribution vs. magnitude along the Galactic plane.
Dashed lines are zero reference lines.
The left and center histograms present the absolute difference
'data minus model' in J-K$_s$ colour and K$_s$
magnitude, and the range between two zero lines corresponds to 70000
mag$^{-1}$deg$^{-2}$. The right ones present the relative
difference in K$_s$ counts($m$ being the average of model counts
for 11.5~$\leq$~K$_s$~$<$~13.5)
and the step between two zero lines corresponds to 400\%.
Star counts are compared in
1.5~$\leq$~J-K$_s$~$<$~3 and 11.5~$\leq$~K$_s$~$<$~13.5.}
\label{hist-ldp}
\end{figure*}

Figure \ref{hist-ldp} presents the histograms of differences 'data minus
model' along the galactic plane.
In order to compare the relative excess at different longitudes
apart from changes in disc density from one field to another,
we computed relative differences $\frac{\mbox{data-model}}{m}$ where $m$ is
the average of simulated counts over the full magnitude range.

These histograms show clearly the existence of an
excess of stars\footnote{It should be noted that
the position in K$_s$ and J-K$_s$ of the overdensity depends on the
extinction in each field} at l$\leq$27$^\circ$ (except for the field at
l=$21^\circ$), contrary to fields at l$\geq$28$^\circ$ where data and model
star counts are similar. 

The observed excess is too large to be explained by a bad
determination of the extinction distributions along the line of sight.
Furthermore, color magnitude
diagrams in the Figure \ref{diag-ldp} show that the giant clumps are rather
well reproduced for most fields, and histograms in J-K$_s$ in the same
Figure have similar shapes : only the heights of the maxima change.
Moreover, data are not always totally complete for K close to 13.5,
especially for the most reddened stars in the fields with highest extinction,
which may attenuate the data excess in these fields.

A possible contamination in the counts by stars of the Scutum Spiral
arm cannot explain the observed overdensity a l$\leq$27$^\circ$ as the impact
of the arm is only expected at the position of the tangential point to the
lane of stars, located at l=33$^\circ$, with a
negligible contribution for l$<$31$^\circ$ (Hammersley et al. 1994).

Except for the discrepancies in the most reddest part of the CMDs
(which corresponds to an excess of stars while the great
discrepancies at low longitudes correspond to a deficit of the model),
comparisons of disc fields have shown that the Besan\c{c}on model
reproduces rather well the galactic disc between l=32$^\circ$ and l=45$^\circ$.
Thus, a bad modeling of disc scale length or equivalent disc scale heights
cannot explain the excess of stars, and changes made by realistic
modifications of them would not be sufficient to remove this excess.

\begin{figure}
\resizebox{\hsize}{!}{\epsfig{file=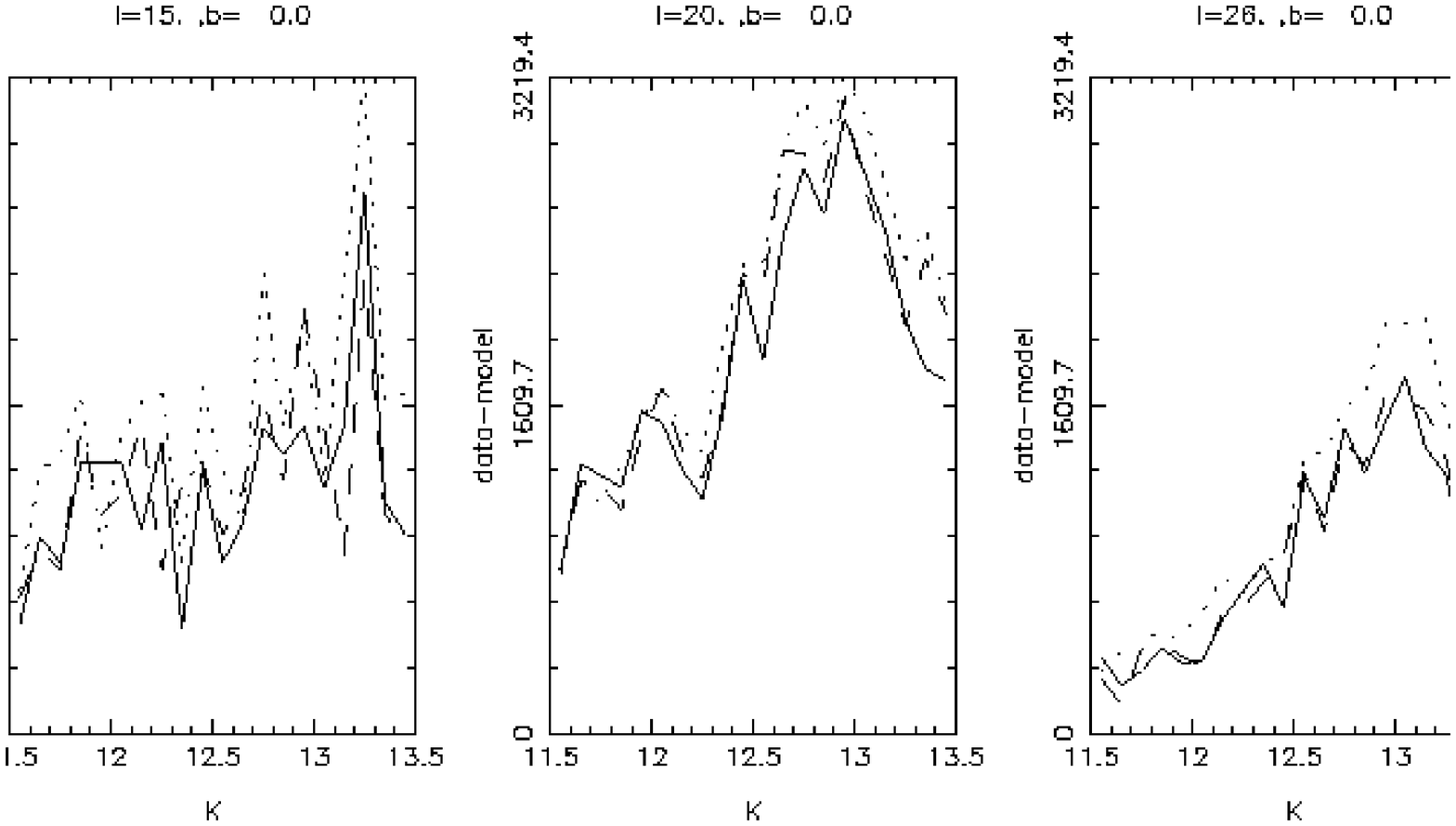}}
\resizebox{\hsize}{!}{\epsfig{file=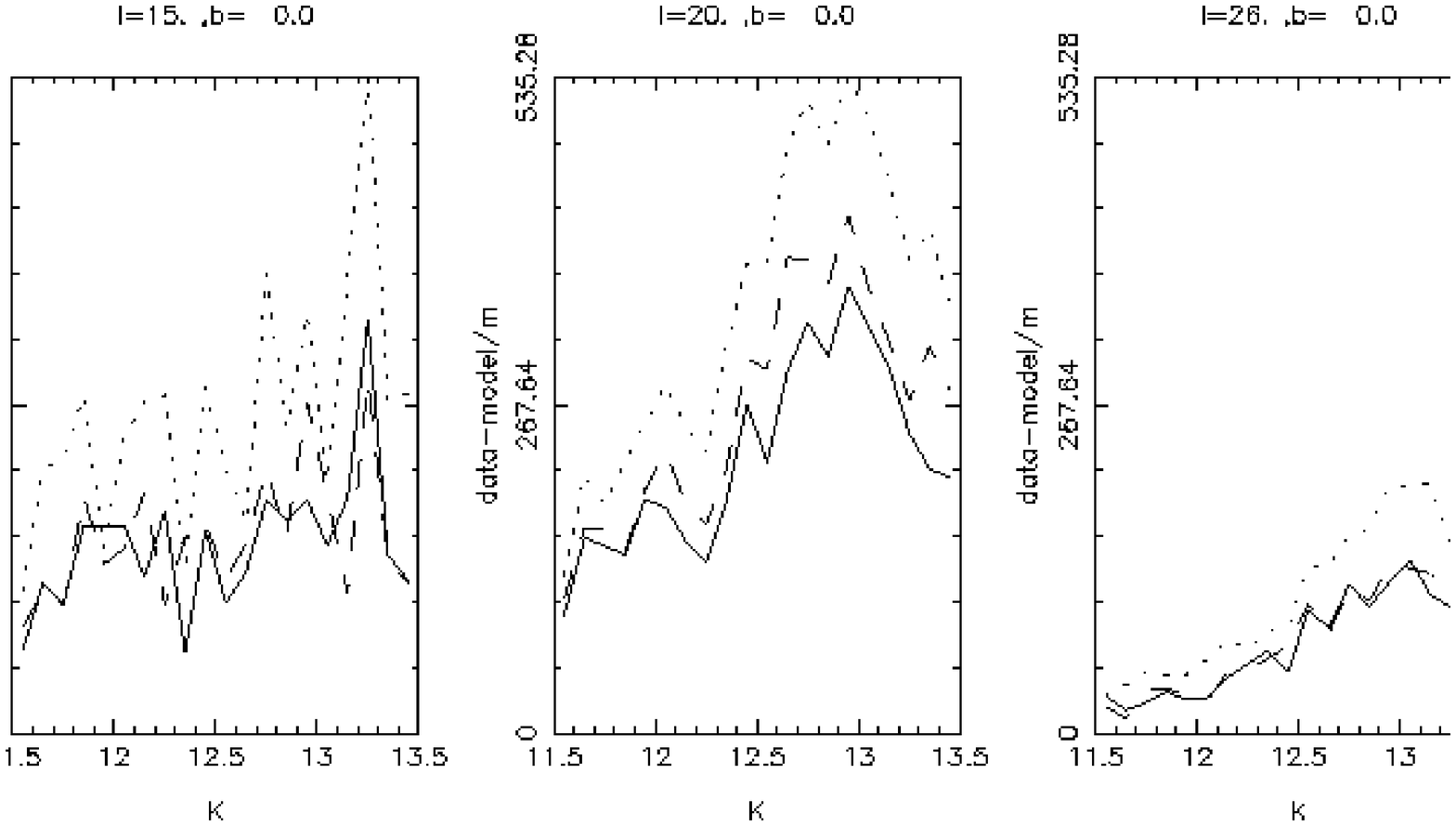}}
\caption{Absolute (upper) and relative (bottom) histograms of difference
'data minus model' in K$_s$ for 3 different hole scale length:  0.5 kpc
(solid lines), 1 kpc (dashed lines) and 1.5 kpc (dotted lines).
Units are
mag$^{-1}$.deg$^{-2}$ for absolute histograms and \% for relative ones.}
\label{rh}
\end{figure}

But this is not the same thing for the hole scale length which, is
not relevant at l$>$32$^\circ$. Fig. \ref{rh} gives the results in 3 fields
b=0$^\circ$ and l=15$^\circ$,20$^\circ$,25$^\circ$ for 3 different hole
scale lengths: 0.5, 1 and 1.5 kpc. Values lower than 0.5 kpc would not
significantly change the histograms, and values greater that 1.5 are not very
realistic. But between these two limits, one can see that the value of the
scale length has a great influence on the excess. Nevertheless, this influence
decreases with longitude and cannot explain all the excess. But it
implies another uncertainty for a quantitative study.

To summarize, the analysis of the histograms of Fig. \ref{hist-ldp}
confirm the existence of an overdensity with respect to the
thin disc predictions. The overdensity begins at
l=27$^\circ$ and extends at least until l=15$^\circ$, but
seems to be inhomogeneous and locally disappears at l=21$^\circ$.

\subsection{Out of the Galactic plane}

\begin{figure*}
\centering
\epsfig{file=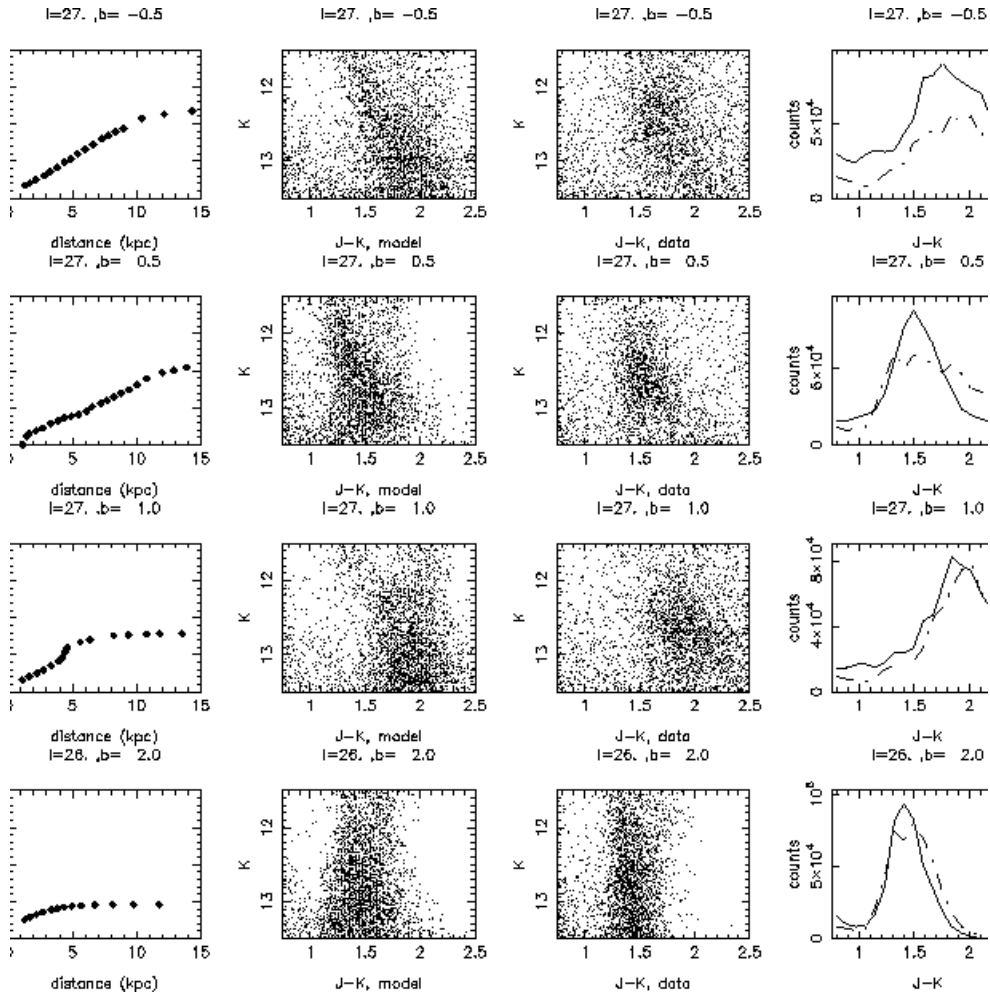,width=13cm}
\caption{Extinction distributions, colour magnitude diagrams and colour
histograms of fields
at b$\neq$0$^\circ$. Explanations of the graphs are given in
Figure \ref{diag-disque}.}
\label{diag-hdp}
\end{figure*}

\begin{figure*}
\centering
\epsfig{file=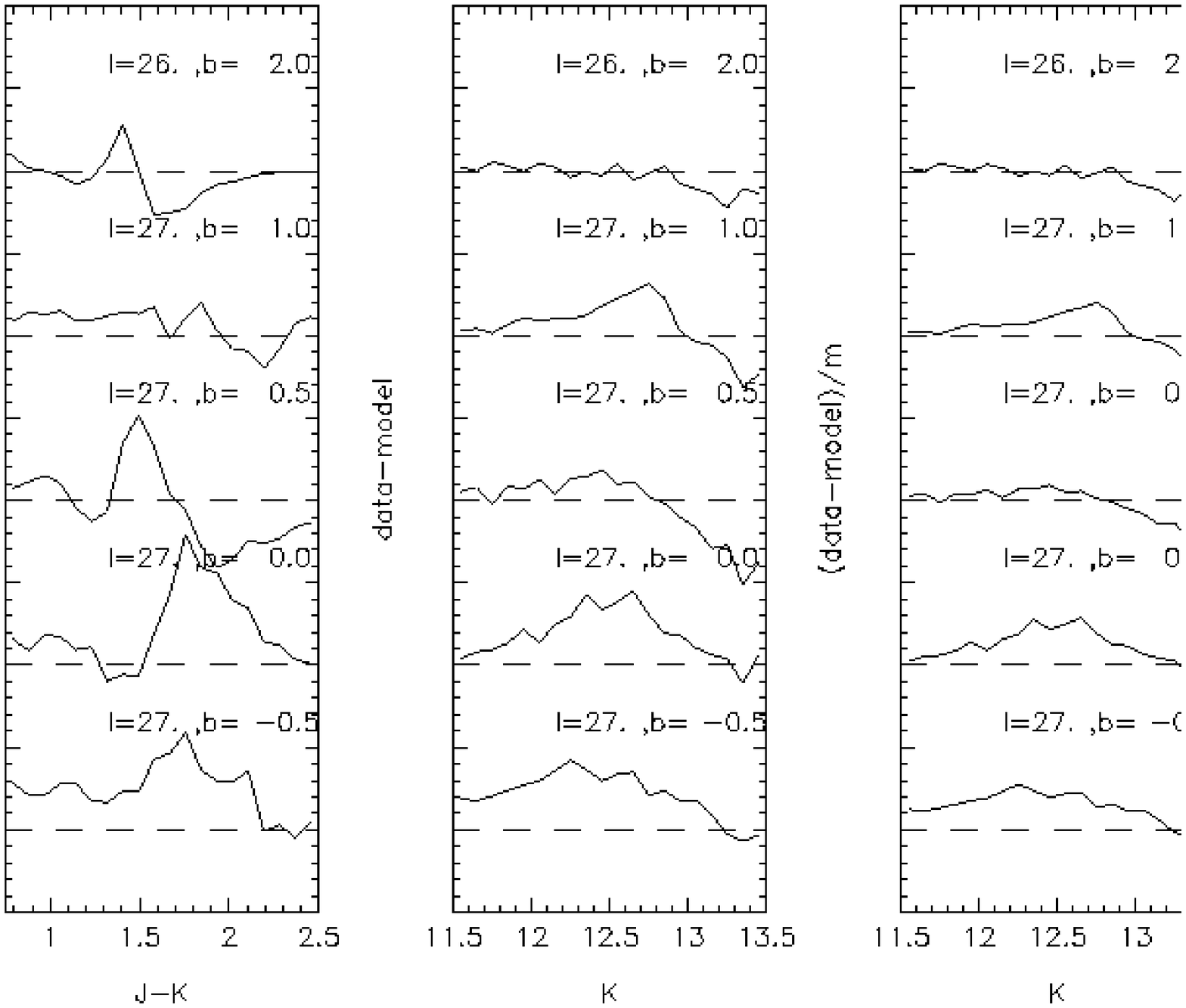,width=13cm}
\caption{Histograms of differences. Explanations of the graphs are given in
Figure \ref{hist-ldp} and the same scales are used.
However we have changed
the cuts on J-K for histograms due to the fact that the reddening is now lower
than in the in-plane regions. The new colour range is
0.75~$\leq$~J-K$_s$~$<$~2.5.}
\label{hist-hdp}
\end{figure*}

We have studied 4 fields out of the galactic plane, roughly at longitudes
coinciding with the end of the overdensity presented in the previous section :
l=27$^\circ$ b=-0.5$^\circ$,+0.5$^\circ$,+1$^\circ$, and l=26$^\circ$
b=+2$^\circ$. The colour-magnitude diagrams and the difference histograms of
theses fields are presented respectively in Figure \ref{diag-hdp} and
\ref{hist-hdp}.

The difficulties in modeling the extinction (even if it is low) in these fields
or to select the giant clump because of the low number of stars, added to the
problem of the Besan\c{c}on model predictions on disc fields out of the plane
make a detailed study of the overdensity difficult out of the galactic plane.

However, Figure \ref{hist-hdp} shows that the excess decreases with
distance from the plane. Indeed, this
excess present at l=27$^\circ$ b=0$^\circ$ still appears at b=-0.5$^\circ$;
It also appears at b=+0.5$^\circ$ but some doubt remains considering the
uncertainties; at b=1$^\circ$ the peak is not significant;
at b=2$^\circ$ the excess has disappeared.

Thus, Figure \ref{hist-hdp} shows that the excess observed at b=0$^\circ$
on different longitudes seems to be confined very close to the galactic plane,
at least at its end (l=26$^\circ$-27$^\circ$).

\subsection{Characterizing the extra density}

\subsubsection{Quantitative considerations}

It is very difficult to quantify the excess density. Indeed,
imperfections in thin disc modeling or extinction modeling and
completeness of data do not prevent the detection of the overdensity but
have an effect on the determination of its characteristics such as density
and position. We try however to extract some information about the
excess density from the comparisons. The most significant parameter
is the star count difference at the maximum
of the peak in K histograms. Table \ref{densites} gives the values for
the 6 fields where the overdensity has been detected and the peak
is rather well defined. Even if values must be taken with great caution,
they show that the excess stars are at least as numerous as the disc ones.

\begin{table*}
{\centering
\begin{tabular}{lcc}
\hline 
\hline
\textbf{Field} & \textbf{Absolute counts (deg$^{-2}$.mag$^{-1}$)}&
\textbf{Relative counts (\%)}\\
\hline
 l=15$^\circ$ b=0$^\circ$&   11335 &      142 \\   
 l=20$^\circ$ b=0$^\circ$&   31734 &      414 \\ 
 l=26$^\circ$ b=0$^\circ$&   18867 &      131 \\
 l=27$^\circ$ b=-0.5$^\circ$& 24571 &      150 \\ 
 l=27$^\circ$ b=0$^\circ$&   28765 &      112 \\
 l=27$^\circ$ b=1$^\circ$&   21998 &      82 \\
\hline
\label{densites}
\end{tabular}\par}
\caption{Star count difference at the maximum of the peak in K histograms.
The second column corresponds to the absolute counts (in number of stars per
square degree per magnitude) $N_{data}-N_{model}$ and the third one
represents the counts (in percentage) relative to the disc
$N_{data}-N_{model}/N_{model}$.}
\end{table*}

\subsubsection{Spatial location of the overdensity}

Hammersley et al. (2000) estimated with CAIN data the position in the
H/J-H CMD of the excess peak at l=27$^\circ$ and deduced, using values of
absolute magnitude and color, a distance of 5.7$\pm$0.7 kpc from the Sun for
excess stars at this field.
Using the distances for the stars from the Besan\c{c}on model
(which has other values for absolute magnitudes and colors),
we have tried to replicate the same determination in the K$_s$ band
making use of the same CAIN database.

3 fields have been used for the determination:
l=26$^\circ$ b=0$^\circ$, l=27$^\circ$ b=0$^\circ$ and
l=27$^\circ$ b=-0.5$^\circ$. In these fields, the excess peak is rather well
defined and extinction seems to be well modeled. The last two fields are not
very reddened, which permits us to avoid problems of incompleteness.
The field at
l=26$^\circ$ is more absorbed but the excess is clearly determined.
Unfortunately, the fields at l=15$^\circ$ and l=20$^\circ$ are too absorbed to
permit a good definition of the the peak excess so they cannot be used to
derive the distance of the stars. Thus, the only fields that can be used
to determine the distance of the excess stars are close to the end of
the overdensity.

\begin{figure}
\resizebox{\hsize}{!}{\epsfig{file=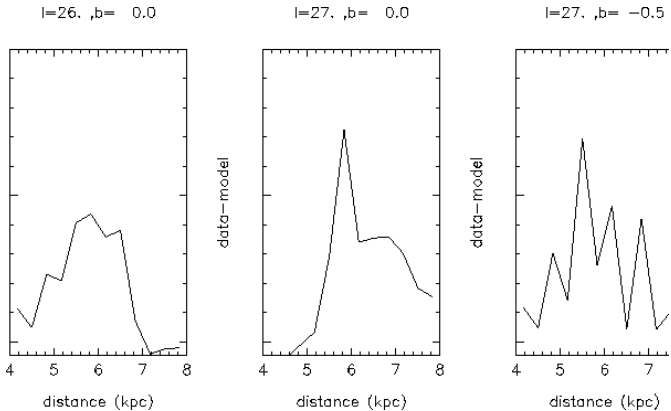,width=13cm}}
\caption{Histograms of distance distribution of the absolute difference
'data minus model'. Distance of data stars were
derived using 12 x 12 bins of colour-magnitude with the following cuts:
12~$\leq$~K$_s$~$<$~13.5 and 1.6~$\leq$~J-K$_s$~$<$~3. for l=26$^\circ$
b=0$^\circ$, and 11.7~$\leq$~K$_s$~$<$~13. and 1.3~$\leq$~J-K$_s$~$<$~2.5 for
l=27$^\circ$ b=0$^\circ$,-0.5$^\circ$.}
\label{distances}
\end{figure}

The following procedure to determine the distance of the excess stars was used:
assuming that these stars are of the same or similar type as the
disc ones, we assigned as distance for each data star in a given
colour-magnitude bin as
the median of the distances of the model stars in the same
bin. Histograms of distance
distributions of the difference 'data minus model' are thus deduced. These
histograms are presented in Figure \ref{distances}.

The distance distribution peaks very clearly at l=27$^\circ$ b=0$^\circ$, with
a maximum at 5.9 kpc. The peak is lower but rather well defined at
l=26$^\circ$ b=0$^\circ$ with the same maximum. At l=27$^\circ$ b=-0.5$^\circ$,
the peak is ever less well defined, with a maximum at 5.5 kpc. We can thus
deduce from these graphs that the stars of the excesses are located at a
distance from the Sun a little lower than 6 kpc. This value confirms
the result given by Hammersley et al. (2000).

\section{Discussion}

The Besan\c{c}on model has shown good performances in producing the observed
NIR counts at intermediate longitudes in the Galactic plane, where the disc is
the major contributor, even if the vertical scalelength of the model
components may have to be fine tuned.

We then confirm the existence of an overdensity between l=20$^\circ$
and l=27$^\circ$ in the Galactic plane, which seems to decrease inversely
to the distance from the galactic plane at its end
(l=26$^\circ$-27$^\circ$), and is located at a distance from the Sun a
little lower than 6 kpc (assuming that they are disc-like stars).

\begin{figure}
\centering
\epsfig{file=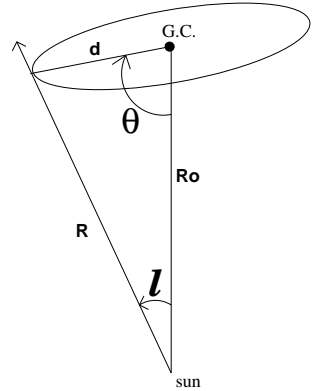,width=4cm}
\caption{Scheme of the galactic in-plane bar}
\label{shema}
\end{figure}

The extension in longitude of the extra density and its 
confinement very close to the galactic plane may suggest an
in-plane bar shape, as is argued in Hammersley et al. (2000) and
L\'opez-Corredoira et al. (2001). Supposing that it does correspond to a bar,
one can estimate its bar angle and its half-length, using the distance of
excess stars from the Sun at its top end.
Neglecting the width of the bar end and taking a distance between the
Sun and the galactic center $R_\odot=8\pm0.5$ kpc, and a distance from the
Sun of bar stars $R=5.9\pm0.5$ kpc, we obtain (see Figure \ref{shema}) that
the bar has a half-length $d=3.9\pm0.4$ kpc and an angle
from the Sun-center direction $\theta=45^\circ\pm9^\circ$.
These results are compatible with the values obtained in
previous works:
$d\approx 4$ kpc and $\theta\approx43\deg$ (Hammersley et al., 2000).
Moreover, other authors have
given similar results for the geometrical parameters of
the Galactic bar by using different methods. Sevenster et al. (1999) extracted
an angle of 44$^\circ$ for the bar through kinematic analysis of OH-IR stars.
Nakai et al. (1992) studied the CO distribution in the inner Galaxy and
suggested a bar near 45$^\circ$, but with non-negligible uncertainties.
More recently, Van Loon et al. (2003) studied luminosity distributions
across the Galactic plane claimed for a bar with an orientation of 40$^\circ$.

Unfortunately, if we can estimate bar parameters assuming as true the
hypothesis that the excess stars belong to a bar population, the fact that the
study was confined at l$\geq$15$^\circ$ because of the completeness of data
makes it impossible to confirm. Even the field
at l=15$^\circ$, very reddened and perhaps a little contaminated by
the outer bulge, must be used with great care. Also, the most inner
fields are dependent on the value of the disc hole scale length.
Furthermore, the field at l=21$^\circ$
b=0$^\circ$, where the excess disappears, shows that the structure has
inhomogeneities. On the other hand, extended star excesses can appear locally,
close to rings for instance, without corresponding to a bar in term of
dynamics.

New observations are needed to settle these uncertainties.
If there is a bar, an excess of stars must exist at negative longitudes
until the other end of the hypothetic bar. With a half-length of $3.9\pm0.4$
kpc and an angle of $45^\circ\pm9^\circ$, the bar far end stars would be
located
at the longitude l=-14$^\circ\pm2^\circ$ and at a distance from the sun
of $11.1\pm1$ kpc. Only much deeper data will allow us to detect them at these
longitudes, and counts will be contaminated by the triaxial bulge in the inner
regions. Moreover, kinematic measurements, at the top end of the hypothetic
bar for instance, are necessary to know whether the extra density corresponds
to a bar in term of dynamics or not. Such a study, using radial velocities,
is being planned.

\section{Acknowledgments}  
The TCS is operated on the island of Tenerife by the Instituto de
Astrof\'{\i}sica de Canarias at the Spanish Observatorio del Teide of the
Instituto de Astrof\'{\i}sica de Canarias.


\begin{thebibliography}{99}

\bibitem[\protect\citeauthoryear{}{}]{} Alves D.R. 2000, ApJ, 539, 732 

\bibitem[ \protect\citeauthoryear{}{}]{}Bahcall, J.N., \& Soneira, R.M.  1980, ApJS, 44, 73

\bibitem[Bienaym\'e et al.(1987)]{modele87}
Bienaym\'e, O., Robin, A.~C., \& Cr\'ez\'e, M. 1987, 
A\&A, 180, 94

\bibitem[Cohen, Hammersley, \& Egan(2000)]{2000AJ....120.3362C} Cohen, M., 
Hammersley, P.\ L., \& Egan, M.\ P.\ 2000, AJ, 120, 3362 

\bibitem[De Jager \& Nieuwenhuijzen (1987)]{dejager} De Jager, C.,
Nieuwenhuijzen, H.  1987, A\&A, 177, 217

\bibitem [\protect\citeauthoryear{}{}]{} Eaton N., Adams D. J, Gilels A. B., 
1984, MNRAS 208, 241 

\bibitem[Einasto(1979)]{Einasto} Einasto, J.  1979,
	IAU Symp. 84, The Large Scale Characteristics of the Galaxy,
	ed. W.B. Burton, p. 451

\bibitem [\protect\citeauthoryear{}{}]{} Epchtein N.  1997, in: The Impact of
Large Scale Near-IR Sky Surveys, F. Garz\'on F., N. Epchtein N., A. Omont,
B. Burton, P. Persi, eds., kluwer, Dordrecht, p.\ 15

\bibitem[ \protect\citeauthoryear{}{}]{} Garz\'on, F., Hammersley, P.L., Mahoney, T., et al.  1993, MNRAS, 264, 773.

\bibitem[Garzon et al.(1997)]{1997ApJ...491L..31G} Garzon, F.,
Lopez-Corredoira, M., Hammersley, P., Mahoney, T.~J., Calbet, X., \&
Beckman, J.~E.\ 1997, ApJ, 491, L31

\bibitem[Garzon et al.(1999)]{} Garzón F. 1999, in:
The Evolution of Galaxies on Cosmological
Timescales, ASP Conf. Ser. 187, ed. J. E. Beckman, \& T. J Mahoney
(Sheridan Books, S. Francisco), 31

\bibitem[\protect\citeauthoryear{}{}]{} Grocholski, A.J., Sarajedini, A. 2002, AJ, 123, 1603 

\bibitem [\protect\citeauthoryear{}{}]{} Hammersley P. L., Garz\'on F.,
Mahoney T., Calbet X.  1994, MNRAS 269, 753

\bibitem [\protect\citeauthoryear{}{}]{} Hammersley, P.L., Garzon, F., Mahoney, T., \& Calbet, X. 1995, MNRAS, 
273 206 

\bibitem[Hammersley et al.(2000)]{2000MNRAS.317L..45H} Hammersley, P.\ L., 
Garz{\' o}n, F., Mahoney, T.\ J., L{\' o}pez-Corredoira, M., \& Torres, M.\ 
A.\ P.\ 2000, MNRAS 317, L45 

\bibitem[Haywood et~al.(1997)]{Misha}
{Haywood}, M., {Robin}, A.~C., \& {Cr\'ez\'e}, M. 1997
A\&A, 320, 440

\bibitem[Lejeune et~al.(1997)]{Lejeune1997}
{Lejeune}, T., {Cuisinier}, F., \& {Buser}, R. 1997
A\&AS, 125, 229

\bibitem[Lejeune et~al.(1998)]{Lejeune1998}
{Lejeune}, T., {Cuisinier}, F., \& {Buser}, R. 1998
A\&A, 130, 65

\bibitem[L{\' o}pez-Corredoira et al.(1999)]{Lopez99}
L\'opez-Corredoira, M., Garz\'on, F., Beckman, J.~E., Mahoney, T.~J.,
Hammersley, P.~L., \& Calbet, X.\ 1999, AJ, 118, 381

\bibitem [\protect\citeauthoryear{}{}]{} L\'opez-Corredoira M.,
Hammersley P. L., Garz\'on F., Simonneau E., Mahoney T. J. 2000, MNRAS
313, 392

\bibitem [ \protect\citeauthoryear{}{}]{} L\'opez-Corredoira, M.,
Hammersley, P.L., Garz{\' o}n, F., Cabrera-Lavers, A., Rodr\'iguez N.,
Schultheis, M., Mahoney T. J. 2001, A\&A, 373, 139

\bibitem[ \protect\citeauthoryear{}{}]{} L\'opez-Corredoira, M., 
Cabrera-Lavers, A., Garz{\' o}n, F., Hammersley, P.L. 2002, A\&A, 394, 883
(L02)

\bibitem[Mathis (1990)]{Mathis}
Mathis, J.S. 1990 ARA\&A 28,37

\bibitem[ \protect\citeauthoryear{}{}]{} Nakai, N.  1992, PASJ, 44, L27

\bibitem[ \protect\citeauthoryear{}{}]{} Paul, E.R. 1993, in The Milky Way Galaxy and Statistical Cosmology, 1890-1924
(Cambridge University Press, Cambridge).

\bibitem[Reyl\'e \& Robin(2001)]{Celine}
Reyl\'e, C., Robin, A.C. 2001
A\&A 373,886

\bibitem[Robin \& Cr\'ez\'e(1986)]{modele86} 
Robin, A.~\& Cr\'ez\'e, M. 1986, A\&A, 157, 71

\bibitem[Robin et~al.(1996)]{Annie96}
Robin, A.C., Haywood, M., Cr\'ez\'e, M., Ojha, D.K., Bienaym\'e, O.  1996
A\&A, 305,125

\bibitem[Robin et~al.(2000)]{Annie00}
Robin, A.C., Reyl\'e, C., Cr\'ez\'e, M. 2000
A\&A, 359,103

\bibitem[Robin et~al.(2002)]{modele2003}
Robin, A.C., Reyl\'e, C., Derri\`ere, S., Picaud, S. 2003
accepted by A\&A

\bibitem[Ruphy et~al.(1996)]{Ruphy}
Ruphy, S., Robin, A.C., Epchtein, N., Copet, E., Bertin, E., Fouqu‰\'e, F.,
Guglielmo, F.  1996
A\&A, 313, 21

\bibitem [\protect\citeauthoryear{}{}]{} Sevenster, M., Saha, P., Calls-Gabaud,
D., Fux, R.  1999, MNRAS 307,584

\bibitem [\protect\citeauthoryear{}{}]{} Skrutskie, M. F., Schneider S. E.,
Stiening R., et al.  1997, in: The Impact of
Large Scale Near-IR Sky Surveys, F. Garz\'on F., N. Epchtein N., A. Omont,
B. Burton, P. Persi, eds., kluwer, Dordrecht, p.\ 25

\bibitem [\protect\citeauthoryear{}{}]{} van Loon, J.T., Gilmore, G.F., Omont, A.,
Blommaert, J.A.D.L., et al. 2003, MNRAS, 338, 857 

\bibitem [\protect\citeauthoryear{}{}]{} Wainscoat R. J., Cohen M., Volk K., 
Walker H. J., Schwartz D. E.  1992, ApJS 83, 111

\end{thebibliography}
\end{document}